\documentclass[twocolumn,prb,nobibnotes]{revtex4}
\usepackage{inputenc}
\inputencoding{ansinew}
\usepackage[T1]{fontenc}
\usepackage{graphicx}
\usepackage[Gray,squaren]{SIunits}
\usepackage{amsmath}
\usepackage{amssymb}

  \usepackage{palatino}
  \usepackage{mathpazo}

\begin{document}
\author{Bo Jakobsen}  
\email{boj@ruc.dk}
\author{Claudio Maggi}
\author{Tage Christensen} 
\author{Jeppe C. Dyre}
\affiliation{DNRF Centre ``Glass and Time'', IMFUFA, Department of Sciences,
Roskilde University, Postbox 260, DK-4000 Roskilde, Denmark}

\date{\today}

\title{Investigation of the shear-mechanical and 
  dielectric relaxation processes in two mono-alcohols close to the
  glass transition}

\begin{abstract}
  Shear-mechanical and dielectric measurements on the two monohydroxy
  (mono-alcohol) molecular glass formers 2-ethyl-1-hexanol and
  2-butanol close to the glass transition temperature are presented.
  The shear-mechanical data are obtained using the piezoelectric
  shear-modulus gauge method covering frequencies from $1\milli\hertz$
  to $10\kilo\hertz$.  The shear-mechanical relaxation spectra show
  two processes, which follow the typical scenario of a structural
  (alpha) relaxation and an additional (Johari-Goldstein) beta
  relaxation.  The dielectric relaxation spectra are dominated by a
  Debye-type peak with an additional non-Debye peak visible. This
  Debye-type relaxation is a common feature peculiar to mono-alcohols.
  The time scale of the non-Debye dielectric relaxation process is
  shown to correspond to the mechanical structural (alpha) relaxation.
  Glass-transition temperatures and fragilities are reported based on
  the mechanical alpha relaxation and the dielectric Debye-type
  process, showing that the two glass-transition temperatures differ
  by approximately $10\kelvin$ and that the fragility based on the
  Debye-type process is a factor of two smaller than the structural
  fragility.  If a mechanical signature of the Debye-type relaxation
  exists in these liquids, its relaxation strength is at most $1\%$
  and $3\%$ of the full relaxation strength of 2-butanol and
  2-ethyl-1-hexanol respectively. These findings support the notion
  that it is the non-Debye dielectric relaxation process that
  corresponds to the structural alpha relaxation in the liquid.
\end{abstract}

\maketitle

\section{Introduction}
\label{sec:introduction}
A class of often investigated glass-forming liquids is the
hydrogen-bonding liquids, among which the alcohols are a much studied
subclass (for a compilation of references to classical results prior
to 1980 see Ref.\ \onlinecite{Bottcher1980}, section IX-c.1).
Alcohols are normally classified into those containing one hydroxyl
group (the mono-alcohols) and those with two or more hydroxyl groups.

During the 1950's it was observed that the main relaxation in
most mono-alcohols, contrary to the main relaxation in other liquids, can
be represented by a single relaxation time --- they follow the Debye
prediction \cite{Davidson1951}.  It was further realized that additional
relaxation processes exist at frequencies above the main
Debye-type relaxation. One additional process is normally observed,
but in some cases two processes can be resolved (see, e.g., Ref.
\onlinecite{Cole1952}). Comparisons between mechanical and dielectric
measurements\cite{Kono1966,Litovitz1963} further showed that when the
main dielectric relaxation is of Debye-type its time scale is separated
from the mechanical time scale, but no explanation was given
for this. It was further discussed to what extent the Debye-type
process corresponds to the mechanical relaxation, as e.g. stated
by Johari and Goldstein\cite{Johari1971} discussing the importance
of mechanical measurements near the glass transition temperature:
``such a study can answer an important question: whether or not the
same molecular motions are involved in the volume, shear, and
dielectric relaxation of H-bonded liquids''. 

During the last decade a number of
studies\cite{Murthy1996,Murthy1996a,Hansen1997,Kudlik1997,Wendt1998,Murthy2002,Wang2004,Wang2005a,Wang2005b,Wang2005,Huth2007,Wang2008,Goresy2008}
(see below for details) have indicated that the low-frequency
Debye-type peak is decoupled form the mechanical relaxation and that
the non-Debye dielectric peak at higher frequencies reflects the
structural alpha relaxation.  In this paper we shall term the two
lowest frequency dielectric relaxations the Debye-type relaxation and
the alpha relaxation, respectively.

This scenario offers an explanation for the earlier observations and it
gives the possibility that the behavior of mono-alcohols follows that
of other glass formers, except for the existence of the Debye-type
dielectric peak.

Two classes of arguments are generally given for this idea: 
Comparisons of time scales/glass-transition temperatures and the lack
of a Debye-type peak in other measurement types. A large number of
experiments and comparisons exists including the following: Comparison
with calorimetric measurements\cite{Murthy1996,Murthy1996a}; Comparison
with photon correlation spectroscopy probing the density-density
correlations\cite{Hansen1997}; Comparison with the time scale found from
viscosity data \cite{Hansen1997,Wang2004}; Analysis of the alpha--beta
relaxation\cite{Kudlik1997}; Solvation dynamics probing mechanical
relaxation of the liquid \cite{Wendt1998}; Dielectric and calorimetric
investigation of mixtures of mono-alcohols with other substances
\cite{Murthy2002,Wang2005a,Wang2005b,Wang2005}; frequency-dependent
specific heat measurements \cite{Huth2007}; systematic comparison 
to DSC calorimetric measurements\cite{Wang2008}; dielectric studies of
mixtures\cite{Goresy2008}.

Except for the early ultrasonic-based
measurements\cite{Lyon1956,Kono1966} no direct comparison exists, to
the best of our knowledge, of the macroscopic mechanical relaxation
spectra and dielectric relaxation spectra of mono-alcohols.

Such measurements directly reveal if the non-Debye relaxation mode
seen in dielectrics is in accordance with the shear-mechanical
structural alpha relaxation.  Shear-mechanical relaxation data are
furthermore generally a good complement to dielectric
data\cite{Jakobsen2005}, and such investigations can also explore to
what extent a shear-mechanical Debye-type relaxation exists.

In this study we present shear-mechanical investigations in the
temperature range down to the glass transition temperature together
with complementary dielectric spectroscopy investigation to allow for
direct comparison.

The two liquids studied are 2-butanol, and 2-ethyl-1-hexanol.  The
reason for not studying simple normal-alcohols like ethanol, is that
such systems easily crystallize. The chosen systems represent two
ways of introducing steric hindrances in the system, hence improving
the glass-forming ability.

Both liquids have been investigated earlier. For early results on
2-butanol see, e.g., Refs. \onlinecite{Dannhauser1955,Murthy1993} and
for 2-ethyl-1-hexanol see, e.g., Refs.
\onlinecite{Wemelle1957,Bondeau1978,Murthy1996a,Murthy2002,Huth2007,Wang2008}.

\begin{figure*}
  \centering
  \includegraphics[width=17cm]{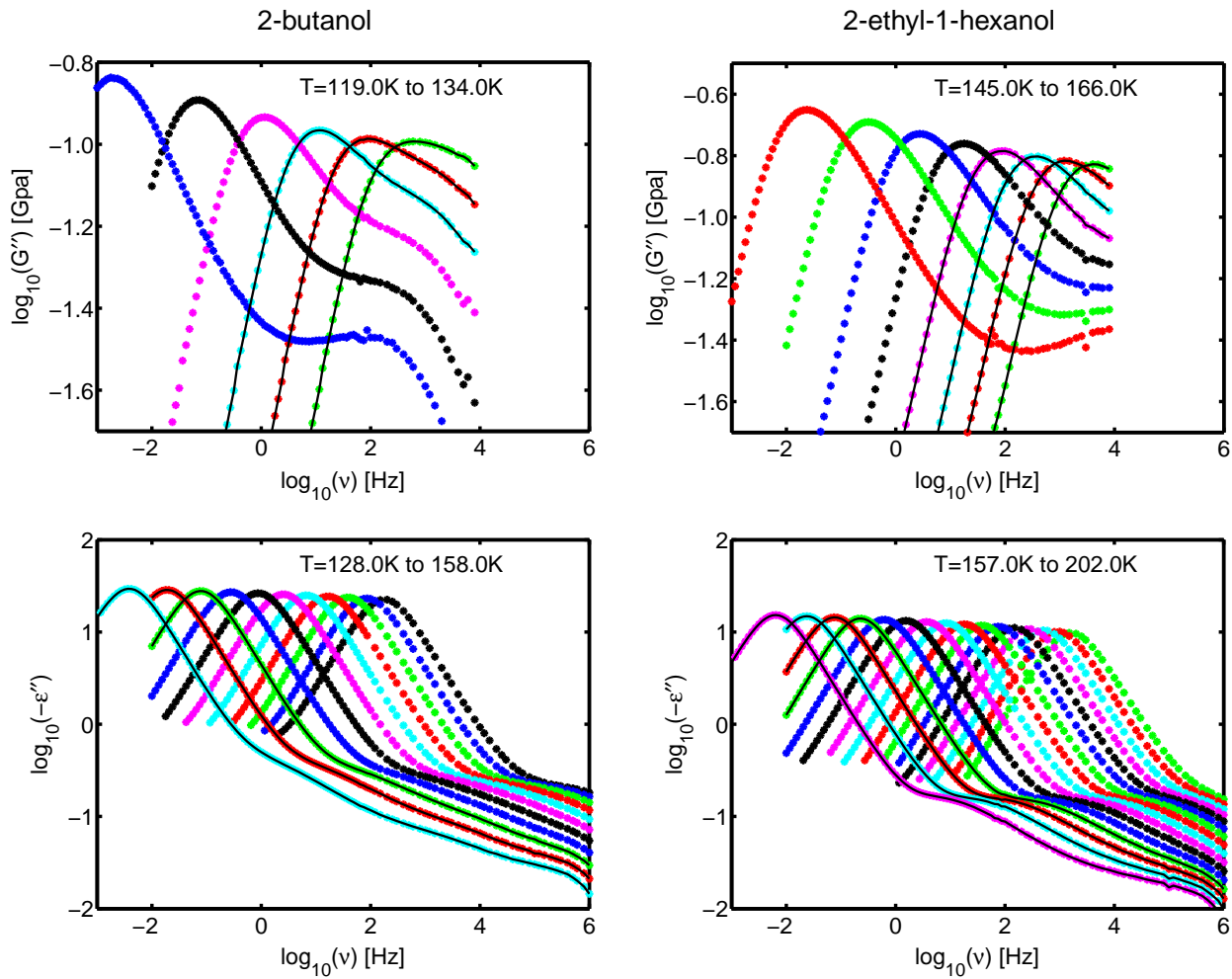}
  \caption{(Color online). Selected dielectric and
      shear-mechanical spectra for the two liquids (the full dataset
      is available online, see Ref.\ \onlinecite{foot3}).  Temperature
      intervals are indicated and the step size is $3\kelvin$.      
      Temperatures which are highlighted by a full line through the
      points exist for both dielectric and shear-mechanical
      measurements, this is only the case in a limited temperature
      interval, where an overlap exists between the high temperature
      shear data and low temperature dielectric data.  \textbf{Top:}
      Shear-mechanical loss as function of frequency.
      \textbf{Bottom:} Dielectric loss as function of frequency.  To
      enhance readability of the figure, the dielectric spectra have
      been truncated at low frequencies, at around the onset of the
      contribution from conduction.}
  \label{fig:loss}
\end{figure*}

\section{Experimental}
\label{sec:experimenaltals}
The measurements were performed using a custom-built
setup\cite{Christensen1995,Igarashi2008a,Igarashi2008b}. The
temperature is controlled by a custom-built cryostat with temperature
fluctuations smaller than $2\milli\kelvin$ (see Ref.\ 
\onlinecite{Igarashi2008a} for details on the cryostat). The same
cryostat was used for all measurements, thus ensuring equal
temperatures and directly comparable results. The electrical signals
were measured using a HP 3458A multimeter in connection with a
custom-built frequency generator in the frequency range of
$10^{-3}\hertz$ -- $10^{2}\hertz$, and a Agilent 4284A LCR-meter in the
frequency range of $10^{2}\hertz$ -- $10^{6}\hertz$ (see Ref.
\onlinecite{Igarashi2008b} for details on the electrical setup).

The shear-mechanical relaxation data were obtained using the
piezoelectric shear modulus gauge (PSG) method\cite{Christensen1995}.
This method has a wide frequency range (up to $10^{-3}$--$10^4$Hz) and
is optimized for measuring moduli in the range of
$\mega\pascal$--$\giga\pascal$, corresponding to typical moduli of
liquids close to the glass transition temperature.  The dielectric
data were obtained using a multilayered gold-plated capacitor with a
empty capacitance of $95\pico\farad$.

2-ethyl-1-hexanol ($\geq99.6\%$, CAS number 104-76-7) and 2-butanol
($99.5\%$, CAS number 78-92-2, racemic mixture) was acquired
from Aldrich and used as received. To ensure that the samples did not
change characteristics (e.g., due to absorption of water) dielectric
measurements were performed on the newly opened bottles and repeated
at the end of the studies. For both liquids the only observable
changes were in the unimportant low-frequency contributions from
conduction.

The raw data\cite{foot3} obtained consist of frequency, $\nu$,
scans of the complex dielectric constant $\epsilon(\nu)$, and the
complex shear modulus $G(\nu)$. Each scan was taken at constant
temperature in thermal equilibrium, stepping down in temperature.

Equilibrium was ensured by repeating some of the frequency scans on
reheating the sample from the lowest temperature. Repetition of
  parts of the shear mechanical measurements showed that the
  uncertainty on the overall absolute level of the shear modulus is
  rather large in the case of 2-ethyl-1-hexanol ($\approx 20\%$), it
  is much better for 2-butanol.  The influence from this experimental
  uncertainty on the position of the loss peaks are however minor (at
  most $\pm 0.1$ decade).

\begin{table}
  \centering
  \caption{Glass transition temperature ($T_g$) and fragility ($m$) 
    for the dielectric Debye-type process
    ($\epsilon_{\text{Debye-type}}$) and mechanical 
    alpha process ($G_{\text{alpha}}$). The glass-transition temperature
    is defined from the loss peak frequencies as $\nu_{lp}(T_g)=10^{-2}\hertz$.}
  \begin{tabular}{l||l|l||l|l}
                      & \multicolumn{2}{c||}{$T_{g}$}&\multicolumn{2}{c}{$m$}\\
                      & $G_{\text{alpha}}$ & $\epsilon_{\text{Debye-type}}$ & $G_{\text{alpha}}$ & $\epsilon_{\text{Debye-type}}$\\ \hline\hline
    2-butanol         
    &$120\kelvin$ \footnote{In accordance with calorimetric $T_g$ of
      $120.3\kelvin$ from Ref.\ \onlinecite{Murthy1993}.} 
    & $130\kelvin$ &$63$ & $29$\\
    2-ethyl-1-hexanol
    &$144\kelvin$ \footnote{In accordance with calorimetric $T_g$ of 
      $145.9\kelvin$ from Ref.\ \onlinecite{Wang2008} and of
      $148.9\kelvin$ from Ref.\  \onlinecite{Murthy1996a}, and
      dielectric alpha-relaxation $T_g$ of $144.0\kelvin$ from Ref.\ \onlinecite{Wang2008}. }
    & $158\kelvin$ \footnote{In accordance with the value of
      $155.0\kelvin$ reported in Ref.\ \onlinecite{Wang2008}, and
      $154.0\kelvin$ reported in Ref.\ \onlinecite{Murthy1996a} (using
      $\nu_{lp}(T_g)=10^{-3}\hertz$ as definition of $T_g$).}
    &$60$ \footnote{Comparable to the value of $70$ reported in Ref.
      \onlinecite{Wang2008} on the basis of dielectric data.}
    & $30$ \footnote{In accordance with the value of $27.0$ reported
      in Ref.\ \onlinecite{Murthy1996a}.}\\
  \end{tabular}
  \label{tab:LiqProp}
\end{table}

\section{Results and discussions}
\label{sec:expresults}
\begin{figure*}
  \centering
 \includegraphics[width=17cm]{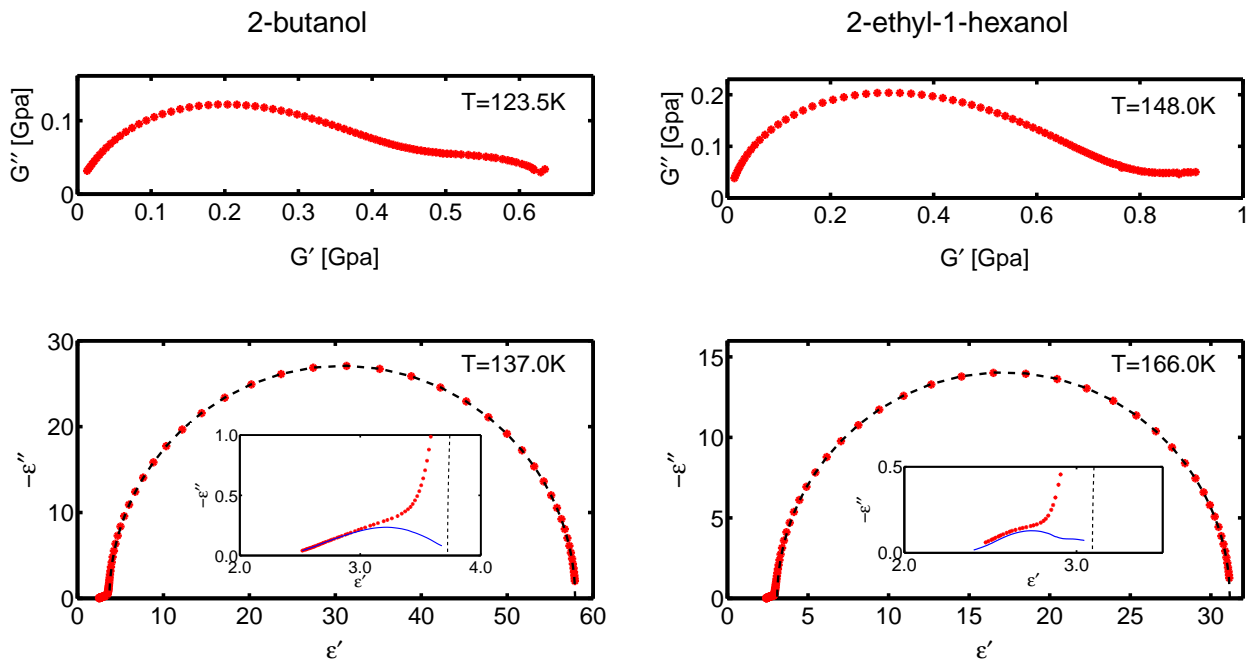}
  \caption{(Color online). Typical dielectric and shear-mechanical
    spectra for the two liquids, represented as Nyquist plots.
    \textbf{Top:} Shear-mechanical spectra.  \textbf{Bottom:}
    Dielectric spectra (the insert shows a zoom on the highfrequency
    foot point). Lines indicate the separation of the dielectric data
    into a Debye-like process (dashed lines) and an alpha relaxation process
    (full lines), see Fig.\ \ref{fig:DebyeSep} for full explanation.}
  \label{fig:nyquist}
\end{figure*}

  A selection of the obtained dielectric spectra is shown in Fig.\
  \ref{fig:loss} represented as the dielectric loss as a function of
  frequency (minus the imaginary part of the complex dielectric
  constant, $-\epsilon(\nu)''$). The dielectric data are furthermore
  illustrated in Fig.\ \ref{fig:nyquist} as a Nyquist plot at a
  representative temperature. The dielectric spectrum follows the
  general pattern for mono-alcohols with a dominant Debye-type
  relaxation, and a minor second relaxation --- the alpha relaxation.

A common way (e.g. Refs.
\onlinecite{Cole1952,Dannhauser1955,Murthy1993,Hansen1997,Wang2004,Goresy2008})
to separate the minor alpha process and possible Johari-Goldstein beta
processes from the Debye-type relaxation process is to assume
additivity of the processes in the dielectric susceptibility
(corresponding to statistical independent dipole-moment fluctuations
of the two processes).  This is either done by fitting a sum of a
Debye function and a Havriliak-Negami function (and possible a
Cole-Cole function for the beta process), or by subtracting the fit of
a Debye function from the raw data (most common in elder studies,
e.g., Ref.\ \onlinecite{Cole1952}).

\begin{figure}
  \centering
 \includegraphics[width=8.5cm]{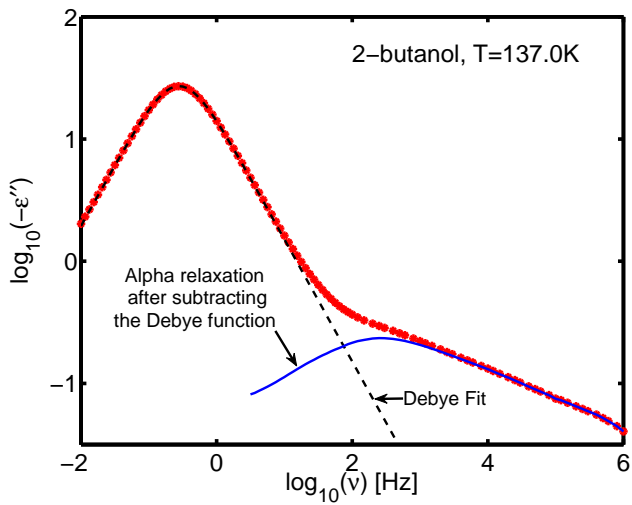}
  \caption{(Color online). Illustration of the procedure used for
      separating the alpha relaxation from the Debye-type relaxation.
      The data shown is the dielectric loss spectrum for 2-butanol at
      $T=137\kelvin$.  Points are measured data, dashed line is a fit
      of the main relaxation to a Debye-function, $\epsilon=\frac{\Delta
        \epsilon_{D}}{1+i\omega\tau}+\epsilon_{{D},\infty}$, and the
      full line the residual after subtracting the Debye-function from
      the measured data (representing the alpha relaxation process).}
  \label{fig:DebyeSep}
\end{figure}

 In this paper we assume additivity of the processes\cite{foot1}
  and subtract the Debye function in order to analyze the residual,
  this procedure is illustrated in Fig.\ \ref{fig:DebyeSep}. The fit
  to the Debye function is also shown in the Nyquist plot in Fig.\
  \ref{fig:nyquist}, illustrating the quality of the fit with respect
  to both real and imaginary part of the dominant dielectric
  relaxation process.

A selection of the shear-mechanical data is shown in Fig.\
\ref{fig:loss} as mechanical loss, $G''(\nu)$, as function of
frequency. Figure \ref{fig:nyquist} furthermore shows the
shear-mechanical relaxation spectra illustrated as a Nyquist plot at a
typical temperature.
The general pattern for liquids close to
the glass transition temperature is observed, with a pronounced
non-Debye alpha relaxation and a smaller Johari-Goldstein beta
relaxation. The beta relaxation is much stronger in the
shear-mechanical relaxation spectrum than in the dielectric spectrum
for these liquids (the existence of a dielectric beta relaxation for
these systems has been reported in the
literature\cite{Murthy1993,Murthy1996a}), consistent with previous
observation on molecular liquids\cite{Jakobsen2005} and the
Gemant-DiMarzio-Bishop model\cite{Niss2005}.

\subsection{Temperature dependence of the dynamics}
\label{sec:loss-peak-positions}
\begin{figure}
  \centering
  \includegraphics[width=8.5cm]{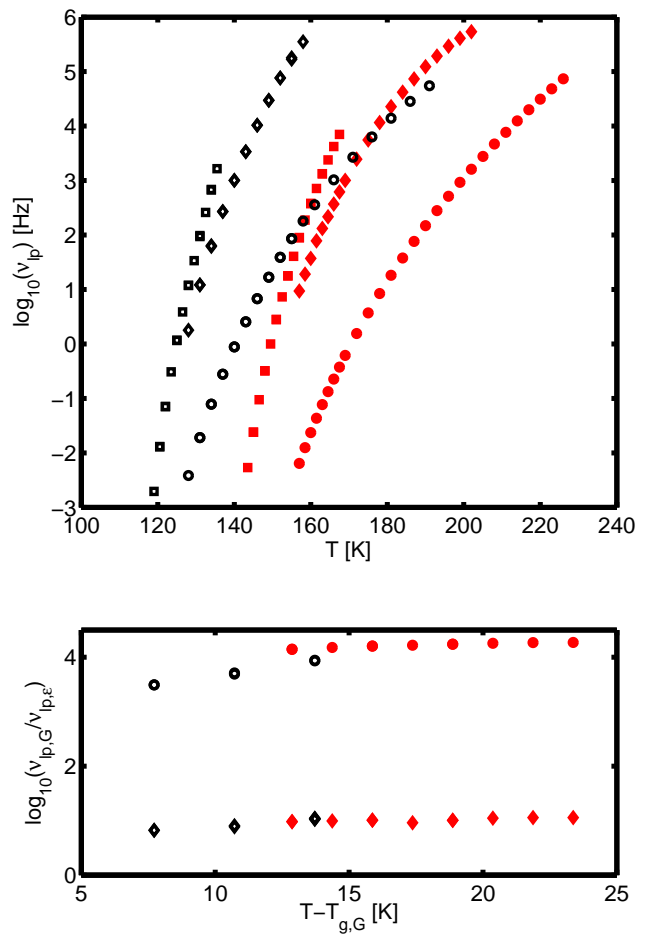}
  \caption{(Color online). \textbf{Top:} Loss-peak positions for the different
    processes evaluated for the two investigated systems (repeated
    measurements exist at some temperatures). 
    \textbf{Bottom:} The decoupling index
    ($\log_{10}(\nu_{lp,G}/\nu_{lp,\epsilon})$) for the dielectric
    alpha and Debye-type process relative to the mechanical alpha
    process. \\
$\square$:
    shear-mechanical alpha relaxation, {\large $\circ$}: dielectric
    Debye-type relaxation, $\lozenge$: dielectric alpha relaxation.
    \textbf{Open symbols:} 2-butanol, \textbf{Closed symbols}:
    2-ethyl-1-hexanol. 
  }
  \label{fig:LossPeakPos}
\end{figure}

To analyze the time scales associated with the observed processes and
their temperature dependencies, the loss-peak frequencies ($\nu_{lp}$)
were determined. These are shown in Fig.\ \ref{fig:LossPeakPos}.  For
the alpha process in the shear-mechanical data and the Debye-type
process in the dielectric data, it was determined directly from the
raw data. For the alpha relaxation in the dielectric data the loss
peak was found after subtracting the Debye function. To ensure
consistency in the analysis the dielectric alpha loss peak was only
calculated at temperatures where the loss-peak of the Debye-type
relaxation was observed.

The dielectric alpha-peak frequency closely follows the peak frequency
of the shear-mechanical alpha process. This is further illustrated in
the lower part of Fig.\ \ref{fig:LossPeakPos} where the ``decoupling''
index (defined as $\log_{10}(\nu_{lp,G}/\nu_{lp,\epsilon})$) is shown
for the dielectric processes (Debye-type and alpha process) relative
to the mechanical alpha process. The mechanical and dielectric
alpha-time scale are separated by approximately one decade in
frequency, whereas the mechanical alpha and the dielectric Debye-type
process are separated by four orders of magnitude in frequency.
The separation between the loss-peak positions can also directly
  be seen on Fig.\ \ref{fig:loss} for the temperatures where both
  shear-mechanical and dielectric data exists (indicated by full
  lines). The separation between the shear-mechanical and dielectric
alpha processes is in agreement with previous comparisons of the
shear-mechanical and dielectric alpha time
scale\cite{Litovitz1963,Kono1966,Menon1994,Christensen1994,Zorn1997,Deegan1999,Schroter2000,Jakobsen2005}.
From Fig.\ \ref{fig:loss} it can be seen that the mechanical beta
  relaxation may influence the loss peak position of the
  shear-mechanical alpha relaxation at high temperatures. From e.g.
  Ref.\ \onlinecite{Jakobsen2005} we know that such influences only
  change the decoupling index between the shear mechanic and
  dielectric alpha relaxations slightly, such an effect can therefor not
  disturb the general observations. It is further noticeable that no
changes can be observed in the temperature dependence of the
shear-mechanical relaxation time around the temperature where the
Debye-type process falls out of equilibrium on the time scale of the
experiment.

The glass transition temperature(s) was determined from the loss peak
frequencies\cite{foot2} as the temperature where
$\nu_{lp}=10^{-2}\hertz$. The numbers for the dielectric Debye-type
relaxation and the shear-mechanical alpha relaxation are given in
table \ref{tab:LiqProp}. The huge separation in time scale between the
two processes results in a separation of $T_g$ of $10\kelvin$ for
2-butanol and $14\kelvin$ for 2-ethyl-1-hexanol.

\begin{figure}
  \centering
  \includegraphics[width=8.5cm]{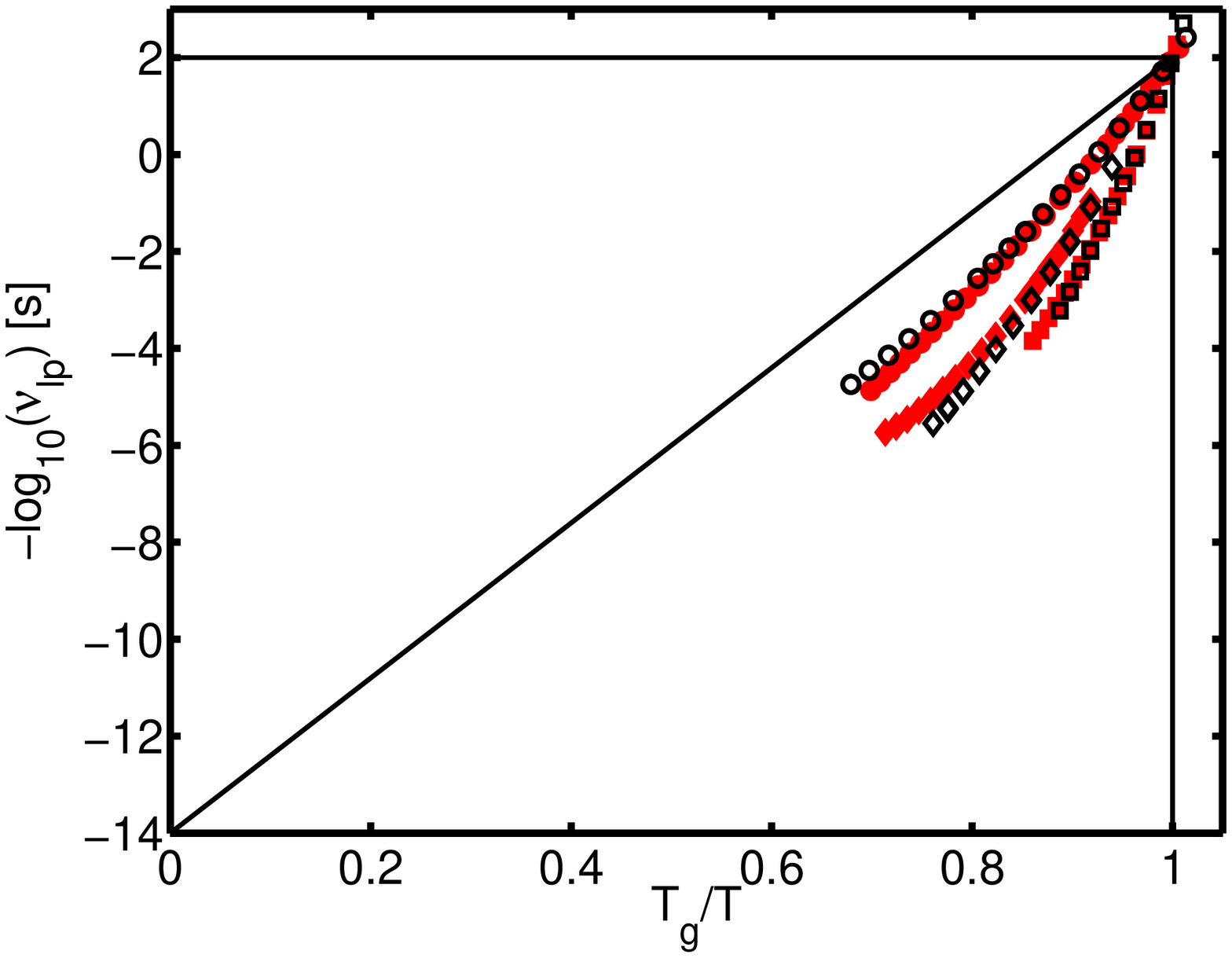}
  \caption{(Color online). Angell fragility plot based on loss peak positions.
    Symbols as on Fig.\ \ref{fig:LossPeakPos}, the vertical and
    horizontal lines defines the glass transition, the diagonal line
    corresponds to Arrhenius behavior.\\
    $T_g$ for the mechanical alpha relaxation and dielectric
    Debye-type relaxation are as given in table \ref{tab:LiqProp}. For
    the dielectric alpha relaxation $T_g$ from the mechanical alpha
    relaxation was used. The reason for this is that the data do not
    allow for direct determination of the $T_g$ for the dielectric
    alpha relaxation without extensively extrapolation, and that the
    two temperature normally not are to different due to the small
    decoupling between the processes.}
  \label{fig:Activation}
\end{figure}

Based on the loss-peak frequencies an Angell plot was constructed as
shown in Fig.\ \ref{fig:Activation}. The two substances show a
remarkable similarity in the temperature dependence of the
characteristic time when plotted this way. The dielectric alpha
process furthermore closely follows the tendencies of the mechanical
alpha process.

The fragility index\cite{Plazek1991,Bohmer1992,Bohmer1993}
$m=\left.\frac{d\log_{10}1/\nu_{lp}}{d T_g/T}\right|_{T=T_g}$ is
reported in table \ref{tab:LiqProp}. A clear difference is seen
between the fragility index if defined from the dielectric Debye-type
process or from the mechanical alpha process. The Debye-type process
leads to a classification of the liquid as much stronger than the
mechanical alpha relaxation.

The difference in temperature dependence of the Debye-type process and
the structural relaxation will lead to a merging of the two processes
at low temperatures, if the trends continuous. This is of course close
to impossible to test experimentally as the relaxation times at such
low temperatures become very long. The possibility of a low
temperature merge of the Debye-type process and the structural alpha
process has been discussed earlier in Ref.\ \onlinecite{Goresy2008} and
\onlinecite{Kalinovskaya2000}. The idea is furthermore supported by the
compilation of data presented in Ref.\ \onlinecite{Wang2004}, the data
generally show a decrease in the difference between the loss peak
frequency of the dielectric Debye-type relaxation and dielectric alpha
relaxation with decreasing temperature.

\subsection{Spectral shape}
\label{sec:peak-shape}
\begin{figure}
  \centering
  \includegraphics[width=8.5cm]{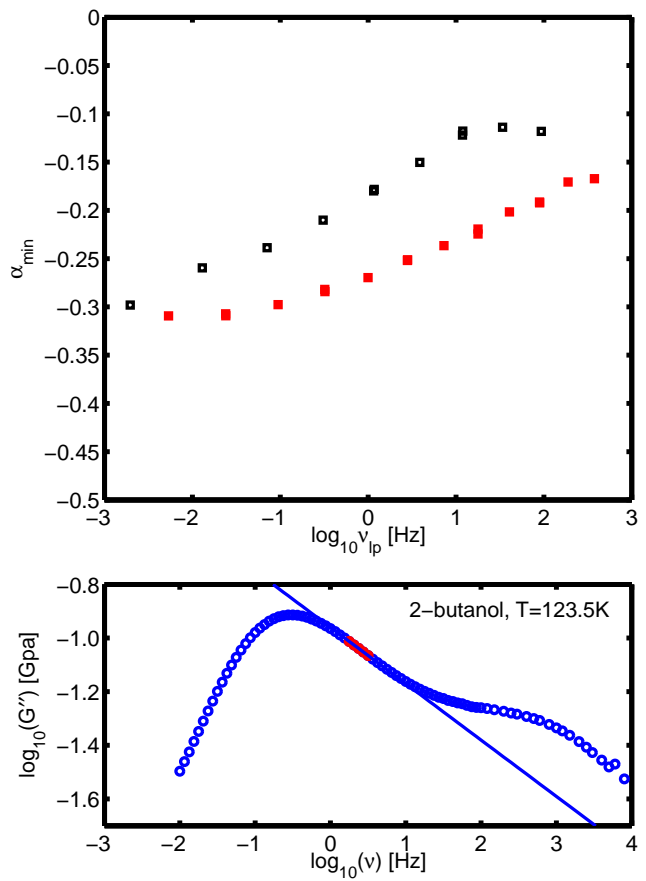}
  \caption{(Color online). \textbf{Top:} Minimum slope of the alpha
    peak ($\alpha_{min}$) in a log-log plot of $G''$ as a function of
    frequency, plotted against the loss-peak frequency (repeated
    measurements exist at some temperatures).  \textbf{Open symbols}:
    2-butanol, \textbf{Closed symbols}: 2-ethyl-1-hexanol.
  \textbf{Bottom:} Illustration of the minimum slope, in a log-log
  plot of $G''$ as a function of frequency. The slope of the
  relaxation curve is found for all frequencies by pointwise
  differentiation, and from this a frequency range of minimum slope is
  determined. The reported minimum slope is an average over the slope
  in this frequency range. In the figure the frequency range of
  minimum slope is marked by closed circles, and the full line goes
  through a central point in this frequency range and has  slope
  equal to the found minimum slope.}
  \label{fig:MinSlope}
\end{figure}

The spectral shape of the shear-mechanical alpha peak, was
characterized by calculation of the minimum slope in a log-log plot as
shown on Fig.\ \ref{fig:MinSlope}.  Similar data have been reported
for shear-mechanical relaxation studies on other systems by our
group\cite{Jakobsen2005,Maggi2008} based on the ideas presented in
Ref.\ \onlinecite{Olsen2001}. Comparing to these previous results it
is observed that the two liquids follow the general trend of liquids
with a mechanical beta relaxation. The minimum slope is in the range
$-0.3$ to $-0.4$ close to $T_g$, still decreasing upon cooling (most
prominent for 2-butanol), possible towards $-0.5$ as conjectured in
Refs.  \onlinecite{Dyre2005,Dyre2006,Dyre2007}.

This shows that the mechanical alpha-relaxation spectra, hence the
mechanical relaxation processes, are similar to what is generally
observed for glass-forming liquids.

\subsection{Limits on a mechanical Debye-type process}
\label{sec:limits-mech-debye}
A small low-frequency peak was observed in the raw data obtained by
the shear-mechanical transducer.  Closer investigations, however,
showed that this was not a mechanical signal of the Debye-type
process, but a ``spillover effect'' of the large dielectric signal.
This effect is caused by wetting of the edges of the piezoceramic
discs in the transducer and the large dielectric strength of the
investigated systems, it is equivalent to sometimes observed conduction
contributions in the raw data.

We can, of course, not entirely exclude that a mechanical signal is
hiding below this dielectric spillover signal, but we are able to put
limits on the maximal relaxation strength. In the case of 2-butanol
the signal was partly eliminated by a correction procedure using data
from a mechanical empty, but still wetted transducer.  From the
resulting shear-mechanical spectra one concludes that a
shear-mechanical relaxation process corresponding to the Debye-type
process in the dielectrics must have a relaxation strength below
$5\mega\pascal$ (corresponding to at most $1\%$ of the full relaxation
strength) if it exists. In the case of 2-ethyl-1-hexanol the raw data
shows that a mechanical Debye-type relaxation process must have a
strength below $30\mega\pascal$ (corresponding to at most $3\%$ of the
full relaxation strength).

\section{Summary and conclusions}
\label{sec:summery}
Two mono-alcohols (2-butanol and 2-ethyl-1-hexanol) were
investigated by conventional dielectric spectroscopy and broadband
shear-mechanical spectroscopy in the temperature range down to the
glass transition temperature.

In the dielectric spectrum a low-frequency Debye-type process is
dominant, as is generally observed for mono-alcohols. The second
relaxation process observed was mathematically separated from the
Debye-type relaxation by assuming additivity of the processes in
the dielectric susceptibility.  Loss-peak positions were found for the
two processes.

Viewed from the shear-mechanical perspective the liquids behave as 
generic glass formers. Besides a clear non-Debye alpha relaxation, a
minor Johari-Goldstein beta relaxation is observed. The loss-peak
positions of the alpha process were determined.

The time scale of the mechanical alpha relaxation is clearly
non-Ahreinius with a fragility index of $\approx 60$ for both liquids.
The Debye-type dielectric relaxation has a much different temperature
dependence, with a fragility index of $\approx 30$ for both liquids.

The time scale of the dielectric alpha relaxation follows closely that
of the mechanical alpha relaxation. The two processes are separated by
roughly one decade in frequency, consistent with what is usually
observed for the separation of mechanical and dielectric alpha
relaxations. The Debye-type process is separated from the mechanical
alpha relaxation by roughly four decades (depending on temperature).
The temperature dependence of the mechanical relaxation time seems to
be unaffected by the falling out of equilibrium of the dielectric
Debye-type process.

The possibility that the Debye-type relaxation process has a
mechanical signature can still not be ruled out, but the present
results show that if it exists one has to use measurement methods
specialized for rather soft systems to look for it. If the Debye-type
process has a mechanical signature, it must have a relaxation strength
below $1\%$ and $3\%$ of the full relaxation strength for 2-butanol
and 2-ethyl-1-hexanol respectively.

These observations supports the existing
idea\cite{Murthy1996,Murthy1996a,Hansen1997,Kudlik1997,Wendt1998,Murthy2002,Wang2004,Wang2005a,Wang2005b,Wang2005,Huth2007,Wang2008,Goresy2008}
that the ``minor'' non-Debye peak observed by dielectric spectroscopy
is the structural alpha relaxation, and that the major Debye-type
relaxation is \textit{something else}.

Any explanation on the dielectric Debye-type relaxation should be able
to explain why no significant signature is observed in either
mechanical or calorimetric studies\cite{Huth2007}.

\section{Acknowledgments}
We are grateful to Niels Boye Olsen for inspiring to this work, and
Kristine Niss for contributing with interesting comments and
questions. This work was supported by the Danish National Research
Foundation's (DNRF) centre for viscous liquid dynamics ``Glass and
Time''.


\begin{thebibliography}{47}
\bibitem[{foo({\natexlab{c}})}]{foot2}
\bibinfo{note}{A loss-peak position of $10^{-2}\hertz$ corresponds to a
  relaxation time of $\approx 10\second$, a factor of $10$ smaller than the
  often used $100\second$. The choice of relaxation time at the glass
  transition temperature is always somewhat arbitrary, and will of course
  influence the reported glass transition temperatures. $10^{-2}\hertz$ has the
  advantage that equilibrium points close to and even below $T_g$ are easily
  accessible, leading to a more price determination of $T_g$.}
\end{thebibliography}
\end{document}